\documentstyle[preprint,prd,aps]{revtex}
\begin{document}

\draft

\title {Double Ernst Solution in Einstein--Kalb--Ramond Theory}

\author{Alfredo Herrera}

\address{Joint Institute for Nuclear Research,\\
Dubna, Moscow Region 141980, RUSSIA, \\
e-mail: alfa@cv.jinr.dubna.su}

\author {\rm and}

\author{Oleg Kechkin}

\address{Nuclear Physics Institute,\\
Moscow State University, \\
Moscow 119899, RUSSIA, \\
e-mail: kechkin@depni.npi.msu.su}

\date{February 1997}

\maketitle

%%%%%%%%%%%%%%%%%%%%%%%%%%%%%%%%%%%%%%%%%%%%%%%%%%%%%%%%%%%%%%%%%%%%%%%%%%%%%%
\draft
%%%%%%%%%%%%%%%%%%%%%%%%%%%%%%%%%%%%%%%%%%%%%%%%%%%%%%%%%%%%%%%%%%%%%%%%%%%%%%
%%%%%%%%%%%%%%%%%%%%%%%%%%%%%%%%%%%%%%%%%%%%%%%%%%%%%%%%%%%%%%%%%%%%%%%%%%%%%%
\begin{abstract}
The K\"ahler formulation of $5$--dimensional
Einstein--Kalb--Ramond (EKR) theory admitting two commuting Killing vectors
is presented. Three different Kramer--Neugebauer--like maps are
established for the $2$--dimensional case. A class of solutions constructed
on the double Ernst one is obtained. It is shown that the double Kerr
solution corresponds to a EKR dipole configuration with horizon.

\end{abstract}
%%%%%%%%%%%%%%%%%%%%%%%%%%%%%%%%%%%%%%%%%%%%%%%%%%%%%%%%%%%%%%%%%%%%%%%%%%%%%%
%%%%%%%%%%%%%%%%%%%%%%%%%%%%%%%%%%%%%%%%%%%%%%%%%%%%%%%%%%%%%%%%%%%%%%%%%%%%%%
\pacs{PACS numbers: 04.20.Jb, 04.50.+h}

\draft

\narrowtext
%%%%%%%%%%%%%%%%%%%%%%%%%%%%%%%%%%%%%%%%%%%%%%%%%%%%%%%%%%%%%%%%%%%%%%%%%%%%%%
%%%%%%%%%%%%%%%%%%%%%%%%%%%%%%%%%%%%%%%%%%%%%%%%%%%%%%%%%%%%%%%%%%%%%%%%%%%%%%
\section{Introduction}
In its low energy limit superstring theory leads to effective Lagrange
systems in which the Einstein action is modified by terms depending
on scalar, vector and tensor fields \cite {ms}--\cite {s4}. 
The Einstein--Kalb--Ramond
(EKR) theory, being a system of this type, takes into account only  the
antisymmetric Kalb--Ramond field, which can be replaced on--shell by
the pseudoscalar Pecci--Quinn axion in four dimensions. The
situation becomes more complicated in the multidimensional case;
the non--trivial properties of this theory were established in 
\cite {s3}--\cite {s2} when dilaton and Abelian gauge fields are present.

In \cite {hk2} it was shown that the ($3 + d$)--dimensional
EKR theory, being reduced to two
dimensions, admits two different $d \times d$--matrix
formulations: the real non--dualized (target space) and the Hermitian
dualized (non--target space) ones; moreover, the Kramer--Neugebauer--like
map between the matrix potentials was established.

In this paper we study the EKR system with $d=2$. It turns out that this
system allows a K\"ahler  representation which formally is defined by
two vacuum Ernst potentials. A discrete transformation between the metric
and Kalb--Ramond degrees of freedom is established; it gives rise to an
alternative $2 \times 2$--matrix formulation of the model. It is shown
that besides of the above mentioned Kramer--Neugebauer map there are two
new  transformations of this type in this case.

In Sec. III the $5$--dimensional line element, explicitly depending on the
Ernst potentials, is presented. After that, we take as starting solution
the Kerr one in order to construct a class of \, $5$--dimensional black
hole solutions; in two of the three analized cases there is a Kalb--Ramond
dipole inside of the horizon.

%%%%%%%%%%%%%%%%%%%%%%%%%%%%%%%%%%%%%%%%%%%%%%%%%%%%%%%%%%%%%%%%%%%%%%%%%%%%%%
%%%%%%%%%%%%%%%%%%%%%%%%%%%%%%%%%%%%%%%%%%%%%%%%%%%%%%%%%%%%%%%%%%%%%%%%%%%%%%
\section{Kramer--Neugebauer Maps}
%%%%%%%%%%%%%%%%%%%%%%%%%%%%%%%%%%%%%%%%%%%%%%%%%%%%%%%%%%%%%%%%%%%%%%%%%%%%%%
%%%%%%%%%%%%%%%%%%%%%%%%%%%%%%%%%%%%%%%%%%%%%%%%%%%%%%%%%%%%%%%%%%%%%%%%%%%%%%
We start from the system with the action
\begin{equation}
{\cal S} = \int d^{5}x {\mid {\cal G} \mid}^{\frac {1}{2}} \left\{ - {\cal R} +
\frac {1}{12} {\cal H}^2 \right\},
\end{equation}
where ${\cal R}$ is the Ricci scalar constructed on the
metric ${\cal G}_{M N}$, $(M = 0,1,...,4)$ and
\begin{equation}
{\cal H}_{MNL} = \partial _{M} {\cal B}_{NL}
+ {\rm cyc.\,\, perms.},
\end{equation}
where ${\cal B}$ is the antisymmetric Kalb--Ramond field and ${\cal H}_{MNL}$
is the non--dualized axion one.

Such a system arises in the frames of the low
energy limit of heterotic string theory. A complete investigation must
include the dilaton and gauge vector fields, but we leave it to be
studied in the near future. Thus, we consider the case when the
vector fields are not present and suggest that the mixings of the metric
(Kaluza--Klein vector fields) and Kalb--Ramond fields vanish:
\begin{equation}
{\cal G}_{\mu, n + 2} = {\cal B}_{\mu, n + 2} = 0,
\end{equation}
where $\mu = 0,1,2$; \, $n = 1,2$.
It is evident that such a restriction does not provide any constraints
on the remainder variables and can be considered as a non--trivial ansatz
for the EKR theory.

After the Kaluza--Klein compactification of two dimensions on a torus,
one obtains the $O(2,2)$--symmetric $\sigma$--model constructed on the
matrix fields $G_{m n}={\cal G}_{m+2,n+2}$ and $B_{m n}={\cal B}_{m+2,n+2}$
coupled to $3$--gravity with the metric
$g_{\mu \nu}={\cal G}_{\mu \nu}$. The effective action of this system is
\cite {ms} and \cite{hk3}
\begin{eqnarray}
^3 S =
\int d^3x {\mid g \mid}^{\frac {1}{2}} \left\{ - ^3 R +
\frac {1}{4} Tr\left[(J^G)^2 - (J^B)^2\right]
\right\},
\end{eqnarray}
where $J^G = \nabla G \, G^{-1}$ and $J^B = \nabla B \, G^{-1}$; $G$ and $B$
being symmetric and antisymmetric $2 \times 2$--matrices, correspondingly.
If we parametrize them as follows:
\begin{eqnarray}
G=p_1 \left (\begin{array}{crc}
p_2^{-1}  & \quad & p_2^{-1}q_2\\
p_2^{-1}q_2 & \quad & p_2 +p_2^{-1}q_2^2\\
\end{array} \right),
\qquad \qquad
B=q_1 \left (\begin{array}{crc}
0  & \quad & -1\\
1 & \quad & 0\\
\end{array} \right);
\end{eqnarray}
the ``material part" of the action (4) takes the form
\begin{eqnarray}
^3 S_m = \frac{1}{2}
\int d^3x {\mid g \mid}^{\frac {1}{2}} \left\{
p_1^{-2}\left[(\nabla p_1)^2 + (\nabla q_1)^2\right] +
p_2^{-2}\left[(\nabla p_2)^2 + (\nabla q_2)^2\right]
\right\},
\end{eqnarray}
which allows us to introduce the independent each other Ernst--like
potentials
\begin{equation}
\epsilon _1 = q_1 + ip_1, \qquad \epsilon _2 = q_2 + ip_2.
\end{equation}
Using these definitions, the action of the model can be rewritten as
\begin{eqnarray}
^3 S =
\int d^3x {\mid g \mid}^{\frac {1}{2}} \left\{ - ^3 R +
2\left(J^{\epsilon _1}J^{\overline\epsilon _1} +
J^{\epsilon _2}J^{\overline\epsilon _2} \right) \right\},
\end{eqnarray}
where $J^{\epsilon _1}=\nabla \epsilon _1
\,(\epsilon _1 - \overline\epsilon _1)^{-1}$
and $J^{\epsilon _2}=\nabla \epsilon _2
\,(\epsilon _2- \overline\epsilon _2)^{-1}$.
Thus, the ``material part"  of the action for the theory under consideration
corresponds to a double Ernst system.

Another $2 \times 2$--matrix representation arises from (5) using
the replacement $p_1 \rightarrow p_2$, \, $q_1 \rightarrow q_2$, which, in
view of (6), provides a discrete symmetry transformation for the action.
This circumstance allows to combine the independent variables $p_1$, $q_1$,
$p_2$ and $q_2$ in the matrices $G'$ and $B'$
\begin{eqnarray}
G'=p_2 \left (\begin{array}{crc}
p_1^{-1}  & \quad & p_1^{-1}q_1\\
p_1^{-1}q_1 & \quad & p_1 +p_1^{-1}q_1^2\\
\end{array} \right),
\qquad \qquad
B'=q_2 \left (\begin{array}{crc}
0  & \quad & -1\\
1 & \quad & 0\\
\end{array} \right);
\end{eqnarray}
in terms of these magnitudes, the action of the system adopts the similar
to (4) form
\begin{eqnarray}
^3 S =
\int d^3x {\mid g \mid}^{\frac {1}{2}} \left\{ - ^3 R +
\frac {1}{4} Tr\left[(J^{G'})^2 - (J^{B'})^2\right]
\right\}.
\end{eqnarray}
In other words, the above mentioned substitution defines the discrete
transformation
\begin{eqnarray}
G \rightarrow G', \quad B \rightarrow B',
\end{eqnarray}
where the matrices $G'$ and $B'$ are considered as the new Kaluza--Klein and
Kalb--Ramond ones; it mixes the gravitational and material degrees
of freedom. It is an analogy of the Bonnor transformation for the stationary
Einstein--Maxwell theory \cite {b} and can be used to generate
new EKR solutions starting with the Kaluza--Klein ones.

In order to establish additional properties of the system under consideration,
we reduce it to two dimensions. In this case, the $3$--metric $h_{ij}$ can
be written in the Lewis--Papapetrou form
\begin{equation}
ds^2 = h_{ij}dx^idx^j = e^{2\gamma}(d\rho ^2+dz^2) - \rho ^2d\tau^2,
\end{equation}
where the function $\gamma$ as well as the ``material" fields are
$\tau$--independent. In the previous work \cite {hk2}, it was shown that
the discussed system allows two different representations in two dimensions:
the target space non--dualized and the non--target space dualized ones.
The target space formulation is given in terms of $G$ and $B$ (or $G'$ and
$B'$); in it the  ``material part" of the $\tau$--independent motion
equations transforms to
\begin{equation}
\nabla (\rho J^B) - \rho J^G J^B = 0,
\end{equation}
\begin{equation}
\nabla (\rho J^G) - \rho (J^B)^2 = 0,
\end{equation}
where $\nabla=\{\partial_{\rho},\partial_{z}\}$. They can be derived from
the $2$--dimensional action
\begin{eqnarray}
^2 S = \frac{1}{4}
\int d\rho dz \rho Tr\left[(J^G)^2 - (J^B)^2\right].
\end{eqnarray}
In the same way we can obtain analogous $2$--dimensional Euler--Lagrange
equations and their corresponding action for $G'$ and $B'$.  Eq. (13),
being rewritten in the form $\nabla[\rho G^{-1}(\nabla B)G^{-1}]=0$,
constitutes the compatibility condition for the relation which defines
the antisymmetric matrix $\Omega$ (by analogy we can define $\Omega '$):
\begin{equation}
\nabla \Omega = \rho G^{-1}(\tilde\nabla B) G^{-1}
\end{equation}
where $\tilde\nabla_{\rho}=\nabla_z$ and $\tilde\nabla_z=-\nabla_{\rho}$
(see \cite {k}). The sets of matrices $(\Omega,G)$ and $(\Omega ',G')$
provide alternative Lagrange
descriptions of the $2$--dimensional theory. In terms of $\Omega$ and $G$
the action of the system adopts the form
\begin{eqnarray}
^2 S = \frac{1}{4}
\int d\rho dz Tr\left[\rho (J^G)^2 + \rho^{-1}(J^{\Omega})^2\right],
\end{eqnarray}
where $J^{\Omega} = G \nabla \Omega$.
For the magnitudes $\Omega '$ and $G'$ we have the same relation. The
Kramer--Neugebauer--like transformations which directly map the ``material"
field equations in terms of $(G,B)$ into the ones in terms of $(G,\Omega)$,
and those given by $(G',B')$ into the ones given by $(G',\Omega ')$ are
\begin{eqnarray}
\left \{ \begin{array}{l}
G \rightarrow \rho G^{-1}\\
B \rightarrow i\Omega  \qquad,\\
\end{array} \right.
\qquad
\left \{ \begin{array}{l}
G' \rightarrow \rho G'^{-1}\\
B' \rightarrow i\Omega ' \qquad;\\
\end{array} \right.
\end{eqnarray}
these transformations are equivalent to the maps
\begin{eqnarray}
\left \{ \begin{array}{l}
p_1 \rightarrow \rho p_1^{-1}\\
q_1 \rightarrow i\omega \qquad,\\
\epsilon _2 \rightarrow -\epsilon _2^{-1}\\
\end{array} \right.
\qquad
\left \{ \begin{array}{l}
p_2 \rightarrow \rho p_2^{-1}\\
q_2 \rightarrow i\omega '\qquad,\\
\epsilon _1 \rightarrow -\epsilon _1^{-1}\\
\end{array} \right.
\end{eqnarray}
where $\omega$ ($\omega '$) is the single non--trivial component of the
matrix $\Omega$ ($\Omega '$), i.e.,
\begin{eqnarray}
\Omega = \omega\left (\begin{array}{crc}
0  & \quad & -1\\
1 & \quad & 0\\
\end{array} \right)
\end{eqnarray}
(a similar relation can be written for $\Omega '$).

The complex matrix transformations (18) have the same form as the well--known
Kramer--Neugebauer one for the pure Einstein theory. These transformations
algebraically map dualized matrix variables into the non--dualized ones.

It is easy to see that Eq. (16) is equivalent to
\begin{equation}
\nabla \omega = \rho p_1^{-2}\tilde\nabla q_1;
\end{equation}
a relation of the same kind  connects $\omega'$ with $p_2$ and $q_2$. Using
such a double non--matrix dualization one can establish a new representation
of the problem. Namely, the equations of motion on the language of $p_1$,
$\omega$, $p_2$ and $\omega '$ correspond to the action
\begin{eqnarray}
^2 S = \frac{1}{2}
\int d\rho dz \left\{
\left[\rho p_1^{-2}(\nabla p_1)^2 - \rho ^{-1}p_1^2(\nabla \omega)^2\right] +
\left[\rho p_2^{-2}(\nabla p_2)^2 - \rho ^{-1}p_2^2(\nabla \omega ')^2\right]
\right\}.
\end{eqnarray}
The map which transforms the motion equations of the action (22) into
the ones of (6) in the two dimensional case, has exactly the
Kramer--Neugebauer \cite {kn} form for the sets of primed and 
non--primed variables:
\begin{eqnarray}
\left \{ \begin{array}{l}
p_1 \rightarrow \rho p_1^{-1},\\
q_1 \rightarrow i\omega, \qquad\\
p_2 \rightarrow \rho p_2^{-1},\\
q_2 \rightarrow i\omega '.\qquad\\
\end{array} \right.
\end{eqnarray}
Thus, 5--dimensional EKR theory, being reduced to two dimensions, allows
three different transformations of Kramer--Neugebauer type.
%%%%%%%%%%%%%%%%%%%%%%%%%%%%%%%%%%%%%%%%%%%%%%%%%%%%%%%%%%%%%%%%%%%%%%%%%%%%%%
%%%%%%%%%%%%%%%%%%%%%%%%%%%%%%%%%%%%%%%%%%%%%%%%%%%%%%%%%%%%%%%%%%%%%%%%%%%%%%
\section{Double Ernst Solution}
%%%%%%%%%%%%%%%%%%%%%%%%%%%%%%%%%%%%%%%%%%%%%%%%%%%%%%%%%%%%%%%%%%%%%%%%%%%%%%
%%%%%%%%%%%%%%%%%%%%%%%%%%%%%%%%%%%%%%%%%%%%%%%%%%%%%%%%%%%%%%%%%%%%%%%%%%%%%%
In this section we construct solutions of the $5$--dimensional EKR theory
using its formal equivalence to the double Ernst system \cite {e}.
Namely, it is
easy to see that the solution of the EKR problem, being reduced to two
dimensions and parametrized by the functions $\epsilon _1$, $\epsilon _2$ and
$\gamma$, can be constructed using the solutions of the double vacuum
Einstein equations written in the Ernst form in terms of
$\epsilon _k$ and $\gamma ^{\epsilon _k}$ ($k = 1, 2$)
\begin{eqnarray}
\nabla (\rho J^{\epsilon _k}) = \rho J^{\epsilon _k}(J^{\epsilon _k} -
J^{\bar \epsilon _k}),
\end{eqnarray}
\begin{eqnarray}
\partial _{z} \gamma ^{\epsilon _k} &=&
\rho \left [(J^{\epsilon _k})_z (J^{\bar \epsilon _k})_{\rho}
+ (J^{\bar \epsilon _k})_z (J^{\epsilon _k})_{\rho}\right ],\\
\partial _{\rho} \gamma ^{\epsilon _k} &=&
\rho \left [{|(J^{\epsilon _k}) _{\rho}}|^2
- {|(J^{\epsilon _k})_z}|^2\right ],
\end{eqnarray}
if one identifies
$\gamma \equiv \gamma ^{\epsilon _1} + \gamma ^{\epsilon _2}$.

Thus, the $5$--dimensional line element for the metric ansatz defined by
(5) and (12)
\begin{equation}
ds_5^2 =  e^{2\gamma}(d\rho ^2+dz^2) - \rho ^2d\tau^2 + G_{m n}dx^m dx^n,
\end{equation}
can be expressed using the parametrization of the matrix $G_{m n}$
and the definition of the functions $\epsilon _1$ and $\epsilon _2$
(see Eq. (7)):
\begin{equation}
ds_5^2 =  e^{2\gamma}(d\rho ^2+dz^2) - \rho ^2d\tau^2 +
\frac{\epsilon _1-\bar \epsilon _1}{\epsilon _2 - \bar \epsilon _2}
{\left| du + \epsilon _2 dv \right|}^2,
\end{equation}
where $u=x^3$ and $v=x^4$. The Kalb--Ramond matrix is given by
\begin{eqnarray}
B = \frac {\epsilon _1 - \bar\epsilon _1}{2i}
\left (\begin{array}{crc}
0  & \quad & -1\\
1 & \quad & 0\\
\end{array} \right)
\end{eqnarray}
and the function $\gamma$ corresponds to the potentials $\epsilon _1$ and
$\epsilon _2$ as it was pointed out below.

The metric (28) algebraically depends on the Ernst potentials, so that the
situation in the EKR theory differs from the one that takes place in the
pure Einstein theory. In latter case, in order to construct the metric, one
must use the function $\omega$, which is related with $q_1$ by the
dualization procedure (21), instead of the imaginary part $q_1$ of the
(single) Ernst potential $\epsilon _1$.

As an example of the developed technique which allows to construct EKR
solutions, we present the solution which arises from the double Kerr one.
The Ernst potentials corresponding to two Kerr solutions with sources in
different points of the symmetry axis are:
\begin{eqnarray}
{\cal E}_k = 1 - \frac {2m_k}{r_k + ia_k\cos\theta_k}.
\end{eqnarray}
where $m_k$ and $a_k$ are constant parameters which define masses and
rotations of the sources of the Kerr field configurations.
Our potentials are
$\epsilon _k = i\bar {\cal E}_k$. The two entered coordinate sets are
connected with the polar system as
\begin{eqnarray}
\rho = [(r_k - m_k)^2 + \sigma _k^2]^{\frac{1}{2}}\sin\theta_k,
\qquad
z = z_k + (r_k - m_k)\cos\theta_k,
\end{eqnarray}
where $z_k$ denote the locations of the sources and
$\sigma _k^2 = m_k^2 - a_k^2$. In this case, for the function
$\gamma _k$ one has:
\begin{equation}
e^{2\gamma _k} = \frac {P_k}{Q_k},
\end{equation}
where $P_k = \Delta_k - a_k^2 \sin^2 \theta _k$, \, $Q_k = \Delta_k +
\sigma _k^2 \sin^2 \theta _k$
and $\Delta _k = r_k^2 - 2m_k r_k + a_k^2$.

Then, the Kalb--Ramond matrix $B_{mn}$ will be defined by the function
\begin{equation}
q_1 = \frac {2m_1 a_1 \cos \theta _1}{\Sigma _1},
\end{equation}
where $\Sigma _k = r_k^2 + a_k^2 \cos^2 \theta _k$. This kind of field
configuration corresponds to a Kalb--Ramond dipole with momentum $m_1 a_1$;
it is located in the point $z_1$.

In the particular case when
$\epsilon _1=i\bar {\cal E}$ and $\epsilon _2=i$, the line element
adopts the form (here indeces are dropped)
\begin{equation}
ds_5^2 = P\left(\frac{dr^2}{\Delta}+d\theta ^2\right) -
\Delta sin^2 \theta d\tau^2 +
\left(\frac{\Delta-a^2sin^2\theta}{\Sigma}\right)
\left( du^2 + dv^2 \right).
\end{equation}
This metric describes the gravitational field originated by the massive
(with mass $m$) Kalb--Ramond dipole which is hidden inside of the
horizon
\begin{equation}
r_h=m+\sqrt{m^2-a^2}.
\end{equation}

The discrete transformation (11), which can be written as
$\epsilon _1 \rightarrow \epsilon _2$, $\epsilon _2 \rightarrow \epsilon _1$,
maps this solution into the one defined
by the potentials $\epsilon _1=i$ and $\epsilon _2=i\bar {\cal E}$.
The corresponding metric is:
\begin{equation}
ds_5^2 = P\left(\frac{dr^2}{\Delta}+d\theta ^2\right) -
\Delta sin^2 \theta d\tau^2 +
\left(\frac{\Sigma}{\Delta-a^2sin^2\theta}\right) {\left| du +
i\left(1 - \frac{2m}{r-ia cos\theta} \right) dv \right|}^2.
\end{equation}
In this case the Kalb--Ramond field is absent, i.e., this solution is
a pure Kaluza--Klein one. Thus, the Kalb--Ramond dipole vanishes,
but the metric (36) acquires an additional degree of freedom (the
non--trivial value of \ ${\cal G}_{uv}$) with respect to (34). This
configuration corresponds to the field of a source with mass $m$
which is located inside of the horizon (35).

The double Kerr solution (28)--(29) can be understood as the ``superposition" of
the two considered cases. The simplest ``superposition" of this kind takes
place when the two Ernst potentials coincide:
$\epsilon _1=\epsilon _2=i\bar {\cal E}$.
Therefore, the $5$--dimensional metric reads
\begin{equation}
ds_5^2 =  \frac{P^2}{Q}\left(\frac{dr^2}{\Delta}+d\theta ^2\right) -
\Delta sin^2 \theta d\tau^2 + {\left| du +
i\left(1 - \frac{2m}{r-ia cos\theta} \right) dv \right|}^2.
\end{equation}
This metric has non--diagonal form and, simultaneously, the
Kalb--Ramond field describes a massive (with mass $m$) dipole; it is
also located inside of the horizon defined by the formula (35).

%%%%%%%%%%%%%%%%%%%%%%%%%%%%%%%%%%%%%%%%%%%%%%%%%%%%%%%%%%%%%%%%%%%%%%%%%%%
%%%%%%%%%%%%%%%%%%%%%%%%%%%%%%%%%%%%%%%%%%%%%%%%%%%%%%%%%%%%%%%%%%%%%%%%%%%%%%%%%
\section{Conclusion}

The K\"ahler formulation of $5$--dimensional EKR theory, admitting two
commuting Killing vectors, is presented. In two dimensions there are
three pairs of descriptions of the theory; each pair contains a target
space (non--dualized) representation and a non--target space (dualized)
one. The corresponding Kramer--Neugebauer--like maps between the dualized
and non--dualized variables are established.

A class of solutions constructed on the double Ernst one is obtained. It
turns out that the $5$--dimensional line element explicitly depends on
the Ernst potentials. The double Kerr solution corresponds to an
asymptotically flat EKR dipole configuration.

%%%%%%%%%%%%%%%%%%%%%%%%%%%%%%%%%%%%%%%%%%%%%%%%%%%%%%%%%%%%%%%%%%%%%%%%%%%%%%
%%%%%%%%%%%%%%%%%%%%%%%%%%%%%%%%%%%%%%%%%%%%%%%%%%%%%%%%%%%%%%%%%%%%%%%%%%%%%%
\acknowledgements
We would like to thank our colleagues from the JINR and NPI
for an encouraging relation to our work. One of the authors (A. H.) would like
to thank CONACYT and SEP for partial financial support.
%%%%%%%%%%%%%%%%%%%%%%%%%%%%%%%%%%%%%%%%%%%%%%%%%%%%%%%%%%%%%%%%%%%%%%%%%%%%%%%
%%%%%%%%%%%%%%%%%%%%%%%%%%%%%%%%%%%%%%%%%%%%%%%%%%%%%%%%%%%%%%%%%%%%%%%%%%%%%%%

\end{document}